\documentclass[12pt,journal,onecolumn,draftclsnofoot]{IEEEtran}
\usepackage{graphicx}
\usepackage{amsmath}
\usepackage{cite}
\usepackage{amssymb}
\usepackage{multirow}
\usepackage{color}
\usepackage{epstopdf}
\usepackage{float}
\usepackage{algorithm} 
\usepackage{algorithmic}
\usepackage{flushend}
\usepackage{xhfill}
\usepackage{bm}
\usepackage{tabularray}
\usepackage{acro}
\usepackage{subcaption}
\usepackage{booktabs}
\usepackage{array} 

\usepackage{acro}

\DeclareAcronym{PASS}{
	short = PASS,
	long  = pinching antenna system
}

\DeclareAcronym{2D-PASS}{
	short = 2D-PASS,
	long  = two-dimensional pinching-antenna system
}

\DeclareAcronym{PA-DM}{
	short = PA-IDM,
	long  = pinching-antenna-enabled index and directional modulation
}

\DeclareAcronym{PSO}{
	short = PSO,
	long  = particle swarm optimization
}

\DeclareAcronym{ISAC}{
	short = ISAC,
	long  = integrated sensing and communications
}

\DeclareAcronym{mmWave}{
	short = mmWave,
	long  = millimeter-wave
}

\DeclareAcronym{IoE}{
	short = IoE,
	long  = Internet-of-Everything
}

\DeclareAcronym{FPA}{
	short = FPA,
	long  = fixed-position antenna
}

\DeclareAcronym{FPAs}{
	short = FPAs,
	long  = fixed-position antennas
}

\DeclareAcronym{UPA}{
	short = UPA,
	long  = uniform planar array
}

\DeclareAcronym{ULA}{
	short = ULA,
	long  = uniform linear array
}

\DeclareAcronym{MINLP}{
	short = MINLP,
	long  = mixed-integer nonlinear programming
}

\DeclareAcronym{MILP}{
	short = MILP,
	long  = mixed-integer linear programming
}

\DeclareAcronym{flops}{
	short = flops,
	long  = floating-point operations
}

\DeclareAcronym{DoF}{
	short = DoF,
	long  = degree of freedom
}

\DeclareAcronym{DoFs}{
	short = DoFs,
	long  = degrees of freedom
}

\DeclareAcronym{RF}{
	short = RF,
	long  = radio frequency
}

\DeclareAcronym{BS}{
	short = BS,
	long  = base station
}

\DeclareAcronym{UEs}{
	short = UEs,
	long  = user equipments
}

\DeclareAcronym{UE}{
	short = UE,
	long  = user equipment
}

\DeclareAcronym{3D}{
	short = 3D,
	long  = three-dimensional
}

\DeclareAcronym{2D}{
	short = 2D,
	long  = two-dimensional
}

\DeclareAcronym{1D}{
	short = 1D,
	long  = one-dimensional
}

\DeclareAcronym{BER}{
	short = BER,
	long  = bit error rate
}

\DeclareAcronym{SNR}{
	short = SNR,
	long  = signal-to-noise ratio
}

\DeclareAcronym{SNRs}{
	short = SNRs,
	long  = signal-to-noise ratios
}

\DeclareAcronym{SINR}{
	short = SINR,
	long  = signal-to-interference-plus-noise ratio
}

\DeclareAcronym{MRC}{
	short = MRC,
	long  = maximum ratio combining
}

\DeclareAcronym{MRT}{
	short = MRT,
	long  = maximum ratio transmission
}

\DeclareAcronym{IM}{
	short = IM,
	long  = index modulation
}

\DeclareAcronym{DM}{
	short = DM,
	long  = directional modulation
}

\DeclareAcronym{MIMO}{
	short = MIMO,
	long  = multiple-input multiple-output
}

\DeclareAcronym{PLS}{
	short = PLS,
	long  = physical layer security
}

\DeclareAcronym{OFDM}{
	short = OFDM,
	long  = orthogonal frequency-division multiplexing
}

\DeclareAcronym{QAM}{
	short = QAM,
	long  = quadrature amplitude modulation
}

\DeclareAcronym{SISO}{
	short = SISO,
	long  = single-input single-output
}

\DeclareAcronym{AWGN}{
	short = AWGN,
	long  = additive white Gaussian noise
}

\DeclareAcronym{CSI}{
	short = CSI,
	long  = channel state information
}

\DeclareAcronym{LoS}{
	short = LoS,
	long  = line-of-sight
}

\DeclareAcronym{NLOS}{
	short = NLOS,
	long  = non-line-of-sight
}

\DeclareAcronym{FFT}{
	short = FFT,
	long  = fast Fourier transform
}

\DeclareAcronym{DFT}{
	short = DFT,
	long  = discrete Fourier transform
}

\DeclareAcronym{ISI}{
	short = ISI,
	long  = inter-symbol interference
}

\DeclareAcronym{ICI}{
	short = ICI,
	long  = inter-carrier interference
}

\DeclareAcronym{ML}{
	short = ML,
	long  = maximum likelihood
}

\DeclareAcronym{NML}{
	short = NML,
	long  = near maximum likelihood
}

\DeclareAcronym{AN}{
	short = AN,
	long  = artificial noise
}

\DeclareAcronym{APM}{
	short = APM,
	long  = amplitude-phase modulation
}
\DeclareAcronym{DNN}{
	short = DNN,
	long  = deep neural network
}

\DeclareAcronym{SVD}{
	short = SVD,
	long  = singular value decomposition
}

\DeclareAcronym{EVD}{
	short = EVD,
	long  = eigenvalue decomposition
}

\DeclareAcronym{6G}{
	short = 6G,
	long  = sixth-generation
}

\DeclareAcronym{5G}{
	short = 5G,
	long  = fifth-generation
}

\DeclareAcronym{XL-MIMO}{
	short = XL-MIMO,
	long  = extremely large-scale multiple-input multiple-output
}

\DeclareAcronym{RIS}{
	short = RIS,
	long  = reconfigurable intelligent surface
}

\DeclareAcronym{PA}{
	short = PA,
	long  = pinching antenna
}

\DeclareAcronym{PAs}{
	short = PAs,
	long  = pinching antennas
}

\DeclareAcronym{DL}{
	short = DL,
	long  = deep learning
}

\DeclareAcronym{FA}{
	short = FA,
	long  = fluid antenna
}

\DeclareAcronym{MA}{
	short = MA,
	long  = movable antenna
}

\DeclareAcronym{MAs}{
	short = MAs,
	long  = movable antennas
}

\DeclareAcronym{THz}{
	short = THz,
	long  = terahertz
}

\DeclareAcronym{NOMA}{
	short = NOMA,
	long  = non-orthogonal multiple access
}

\DeclareAcronym{URLLC}{
	short = URLLC,
	long  = ultra-reliable low-latency communication
}

\DeclareAcronym{mMTC}{
	short = mMTC,
	long  = massive machine-type communication
}

\DeclareAcronym{eMBB}{
	short = eMBB,
	long  = enhanced mobile broadband
}

\DeclareAcronym{LDPC}{
	short = LDPC,
	long  = low-density parity-check
}

\DeclareAcronym{Polar}{
	short = Polar,
	long  = polar code
}

\DeclareAcronym{CRC}{
	short = CRC,
	long  = cyclic redundancy check
}

\DeclareAcronym{FEC}{
	short = FEC,
	long  = forward error correction
}

\DeclareAcronym{HARQ}{
	short = HARQ,
	long  = hybrid automatic repeat request
}

\DeclareAcronym{SC}{
	short = SC,
	long  = successive cancellation
}

\DeclareAcronym{DPC}{
	short = DPC,
	long  = dirty paper coding
}

\DeclareAcronym{CSI-RS}{
	short = CSI-RS,
	long  = channel state information reference signal
}

\DeclareAcronym{PRB}{
	short = PRB,
	long  = physical resource block
}

\DeclareAcronym{DMRS}{
	short = DMRS,
	long  = demodulation reference signal
}

\DeclareAcronym{PAPR}{
	short = PAPR,
	long  = peak-to-average power ratio
}

\DeclareAcronym{IUI}{
	short = IUI,
	long  = inter-user interference
}

\DeclareAcronym{ZF}{
	short = ZF,
	long  = zero forcing
}

\DeclareAcronym{MMSE}{
	short = MMSE,
	long  = minimum mean square error
}

\DeclareAcronym{RZF}{
	short = RZF,
	long  = regularized zero forcing
}

\DeclareAcronym{BD}{
	short = BD,
	long  = block diagonalization
}

\DeclareAcronym{THP}{
	short = THP,
	long  = Tomlinson-Harashima precoding
}

\DeclareAcronym{SLNR}{
	short = SLNR,
	long  = signal-to-leakage-and-noise ratio
}

 \IEEEoverridecommandlockouts
 \def\BibTeX{{\rm B\kern-.05em{\sc i\kern-.025em b}\kern-.08em
 		T\kern-.1667em\lower.7ex\hbox{E}\kern-.125emX}}
\begin{document}

%
\title{Two-Dimensional Pinching-Antenna Systems: Modeling and Beamforming Design}

\author
{
	Yuan Zhong,~\IEEEmembership{}
	Yue Xiao,~\IEEEmembership{} 
	Yijia Li,~\IEEEmembership{} 
	Hao Chen,~\IEEEmembership{} 
	Xianfu Lei, ~\IEEEmembership{}
	and
	Pingzhi Fan ~\IEEEmembership{}
	\thanks{
%
		Y. Zhong, Y. Xiao, Y. Li and H. Chen are with the National Key Laboratory of Wireless Communications, University of Electronic Science and Technology of China (UESTC), Chengdu 611731, China (e-mail: yuanzhong@std.uestc.edu.cn;  xiaoyue@uestc.edu.cn; nikoeric@foxmail.com; nclchenhao@std.uestc.edu.cn).	

		X. Lei and P. Fan are with the School of Information Science and Technology, Southwest Jiaotong University, Chengdu 610031, China (e-mail: xflei@swjtu.edu.cn; pzfan@swjtu.edu.cn).
%
	}
}

\maketitle

\begin{abstract}
	Recently, the pinching-antenna system (PASS) has emerged as a promising architecture owing to its ability to reconfigure large-scale path loss and signal phase by activating radiation points along a dielectric waveguide. 
	However, existing studies mainly focus on line-shaped PASS architectures, whose limited spatial flexibility constrains their applicability in multiuser and indoor scenarios. 
	In this paper, we propose a novel two-dimensional (2D) pinching-antenna system (2D-PASS) that extends the conventional line-shaped structure into a continuous dielectric waveguide plane, thereby forming a reconfigurable radiating plane capable of dynamic beam adaptation across a 2D spatial domain. 
	An optimization framework is developed to maximize the minimum received signal-to-noise ratio (SNR) among user equipments (UEs) by adaptively adjusting the spatial configuration of pinching antennas (PAs), serving as an analog beamforming mechanism for dynamic spatial control. 
	For the continuous-position scenario, a particle swarm optimization (PSO)-based algorithm is proposed to efficiently explore the nonconvex search space, while a discrete variant is introduced to accommodate practical hardware constraints with limited PA placement resolution. 
	Simulation results demonstrate that the proposed 2D-PASS substantially improves the minimum SNR compared with conventional line-shaped PASS and fixed-position antenna (FPA) benchmarks, while maintaining robustness under varying user distributions and distances.
\end{abstract}


\IEEEpeerreviewmaketitle

\section{Introduction}
The advent of the \ac{6G} communication era is envisioned to enable an ultra-dense, intelligent, and seamlessly connected network fabric capable of supporting diverse emerging applications such as holographic communications, wireless extended reality, and industrial \ac{IoE} systems~\cite{6G,6G2,IoE_6G}.
To meet the stringent requirements for ultra-high throughput, ultra-low latency, and high reliability, future wireless systems are expected to operate at \ac{mmWave} and \ac{THz} frequency bands, where abundant spectrum resources are available~\cite{MillimeterWaveandTerahertz,terahertz}. 
However, propagation characteristics of high-frequency signals cause severe path loss and frequent \ac{LoS} blockage, resulting in distance-dependent channel power attenuation and restricted coverage~\cite{terahertz2}. 
Therefore, enhancing the spatial directivity and controllability of wireless propagation has become a fundamental challenge for realizing high-capacity and energy-efficient \ac{6G} networks.

In response to this challenge, various reconfigurable and adaptive antenna architectures have been proposed as key physical-layer enablers for future wireless systems. 
Representative examples include \ac{RIS}~\cite{RIS2}, \ac{FA}~\cite{FA1,RIS_FA,FA6}, and \ac{MA} systems~\cite{MA1,MA2,MA3}. 
Specifically, \acp{RIS} employ programmable metasurfaces to passively manipulate electromagnetic (EM) waves, thereby enhancing coverage and spectral efficiency. 
However, their reflective double-hop transmission path often results in reduced end-to-end array gain and increased channel estimation complexity~\cite{FARISBFJ}. 
\ac{FA} systems, in contrast, utilize liquid-metal radiators that can be reconfigured to different spatial positions, providing high adaptability in dynamic environments~\cite{FA2,FA4}. 
However, the discrete port-based channel model adopted in~\cite{FA5} cannot fully characterize or exploit the continuous spatial variations of the wireless channel, thereby limiting its ability to capture fine-grained propagation behavior.
More recently, \ac{MA} systems have been proposed, where each antenna element can move within a confined region to reshape the instantaneous channel condition and exploit spatial channel diversity~\cite{MA4,MA5}. 
Although such designs effectively enhance spatial \ac{DoFs}, their mobility range is physically limited, and they can only adjust the small-scale fading characteristics of the channel rather than mitigate large-scale path loss~\cite{discrete_MA,PA_perspective}.

To overcome these limitations, the \ac{PASS} has been proposed as a novel paradigm for reconfigurable architectures~\cite{PA_perspective,PASS,PASS1}. 
Instead of mechanically moving the entire antenna element, \ac{PASS} employs a dielectric waveguide supporting guided-wave propagation, where localized radiation points are generated by placing small dielectric pinches (e.g., plastic particles) at desired positions along the waveguide~\cite{PASS2}. 
This mechanism perturbs the local field distribution and enables controllable radiation without requiring multiple \ac{RF} chains. 
Building upon this foundation, several pioneering studies have explored \ac{PASS} in various communication contexts. 
For instance, downlink beamforming models were proposed in~\cite{PASS_beamforming,PA_DL_RateMax,PASS_DL2} to jointly optimize \ac{PA} positions and digital precoders, achieving significant rate improvements over fixed arrays. 
The uplink extensions in~\cite{PA_UL_RateMax,PASS_UL,PASS_UL2} jointly optimized transmit power and \ac{PA} locations for rate maximization. 
Moreover, the works in~\cite{PA_IDM,PASS_PLS1,PASS_PLS2} investigated the integration of \ac{PASS} with physical-layer security techniques, demonstrating improved confidentiality through dynamic reconfiguration of radiating positions. 
In parallel, the authors of~\cite{PASS_DL,PA_multiwaveguide,PA_multiwaveguide2} examined multi-waveguide \ac{PASS} architectures enabling joint digital and analog beamforming to enhance multiuser coverage and suppress interference. 
A \ac{NOMA}-assisted \ac{PASS} design was proposed in~\cite{PA_NOMA,PA_NOMA2,PA_NOMA3}, where multiple \acp{PA} were selectively activated via a matching-based algorithm to maximize system throughput with reduced complexity. 
Furthermore, the work in~\cite{PA_beamtraining} developed an efficient beam-training protocol that sequentially activates \acp{PA} along the waveguide for rapid alignment with minimal training overhead. 
To improve scalability, the studies in~\cite{PA_ISAC,PA_ISAC3,PA_ISAC4} introduced a pinching-antenna framework for \ac{ISAC}, jointly optimizing transmit and receive \acp{PA} to balance communication performance and sensing precision. 
Finally, the authors of~\cite{PA_channelestimation} addressed channel estimation challenges for \ac{PASS} and proposed two \ac{DL}-based estimators capable of reconstructing high-dimensional channel coefficients from low-dimensional pilot inputs, thereby improving accuracy and scalability. 
Despite these advances, most existing studies on \ac{PASS} remain limited to line-shaped implementations, where \acp{PA} are arranged along a single waveguide. 
Such an architecture inherently restricts spatial flexibility and constrains the achievable beamforming gain in multiuser or indoor environments.

Motivated by these observations, this paper introduces a \ac{2D-PASS}, which extends the reconfigurability of conventional \ac{PASS} into the planar domain. 
By employing a single integrated dielectric waveguide that continuously extends over a \ac{2D} plane, the proposed \ac{2D-PASS} forms a reconfigurable antenna plane capable of dynamically adjusting its active radiation points along both the $x$- and $y$-axes, thereby realizing analog beamforming with fine spatial granularity. 
This design not only enhances spatial \ac{DoFs} for beam pattern control but also provides superior adaptability for \ac{UE}-centric coverage, particularly in multiuser or indoor scenarios with severe path-loss disparities. 
Moreover, the planar structure naturally supports wall-mounted or ceiling-mounted deployments, offering a cost-effective and scalable solution for next-generation access points.

However, optimizing \ac{PA} positions over a continuous \ac{2D} region yields a highly nonconvex problem. 
The nonlinear coupling between antenna positions and received signal power substantially constrains the performance of gradient-based optimization methods, reducing their ability to reach satisfactory solutions.
To address this, a \ac{PSO} framework is developed, which jointly exploits local exploration and global collaboration to efficiently approach near-optimal configurations without gradient computation. 
Furthermore, to accommodate practical hardware constraints, a discrete-position formulation is introduced where quantized \ac{PA} activations approximate the continuous plane, striking a balance between performance and implementation complexity.

In general, the key contributions of this paper are summarized as follows:
\begin{itemize}
	\item We propose a \ac{2D} extension of the conventional line-shaped \ac{PASS}, in which a single integrated dielectric waveguide forms a continuous planar structure. This architecture significantly increases the spatial \ac{DoFs} and enables highly adaptive analog beamforming for dynamic \ac{UE}-centric transmission.
	\item We establish a comprehensive \ac{3D} geometric model for the proposed \ac{2D-PASS} broadcast system, which incorporates both free-space propagation and in-waveguide phase delay to accurately characterize its electromagnetic behavior.
	\item We develop a \ac{PSO}-based beamforming optimization algorithm for the continuous-position case to maximize the minimum \ac{SNR} among all \acp{UE} under spatial constraints. For the discrete case, we reformulate the position selection task into a \ac{MILP} problem, which can be solved to global optimality using standard solvers.
	\item We demonstrate through simulations that the proposed \ac{2D-PASS} substantially outperforms conventional \ac{PASS} and \ac{FPA} benchmarks in terms of minimum \ac{SNR} and robustness to varying \ac{UE} distributions and distances. Moreover, we verify that with a practically feasible quantization step, the discrete \ac{2D-PASS} achieves performance close to its continuous counterpart with marginal loss, confirming its practical viability.
\end{itemize}

The remainder of this paper is organized as follows. 
Section~\ref{sec:system_model} introduces the system model of the proposed \ac{2D-PASS}. 
Sections~\ref{sec:continuous} and~\ref{sec:discrete} detail the continuous and discrete positioning algorithms, respectively. 
Simulation results are presented in Section~\ref{sec:simulation}, and concluding remarks are drawn in Section~\ref{sec:conclusion}.

\emph{Notation:} Bold lowercase and uppercase letters denote vectors and matrices, respectively.
The operator $(\cdot)^{\mathrm{T}}$ represents the transpose, and $(\cdot)^{*}$ denotes the complex conjugate.
The Euclidean distance between two coordinates ${\bm{\psi}}_i$ and ${\bm{\psi}}_j$ is given by $|{\bm{\psi}}_i - {\bm{\psi}}_j|$.
Moreover, the notation $\mathcal{U}[a,b]$ represents a uniform distribution over the interval $[a,b]$.

\section{System Model}\label{sec:system_model}
\subsection{System Description and Channel Model}
\begin{figure*}[tb]
	\centering
	\begin{subfigure}[t]{0.5\linewidth}
		\centering
		\includegraphics[width=1.02\linewidth]{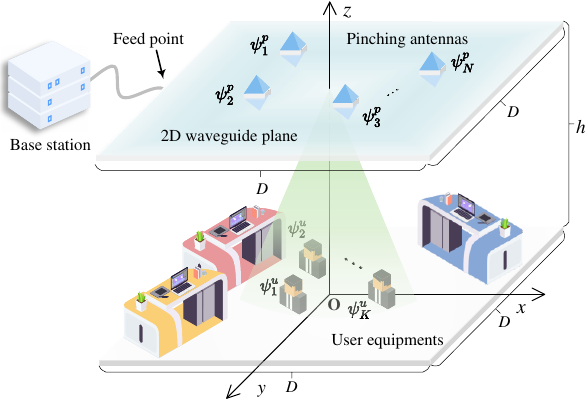}
		\caption{}
		\label{system_model}
	\end{subfigure}%
	\hspace{0em}
	\begin{subfigure}[t]{0.48\linewidth}
		\centering
		\includegraphics[width=1.055\linewidth]{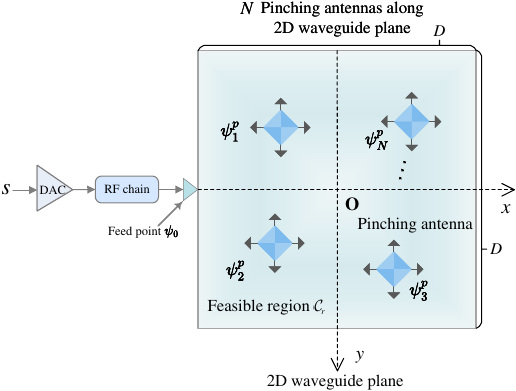}
		\caption{}
		\label{2D_PA}
	\end{subfigure}%
	\caption{System model of the proposed \ac{2D-PASS}: 
		(a) Downlink broadcast beamforming scenario in an indoor environment; 
		(b) Illustration of the \ac{2D} dielectric waveguide plane with $N$ pinching antennas and the feasible placement region.}
	\label{fig:system_model}
\end{figure*}

As illustrated in Fig.~\ref{fig:system_model}(a), we consider a downlink broadcast transmission system based on the proposed \ac{2D-PASS}, where a single \ac{BS} equipped with a \ac{2D} dielectric waveguide simultaneously serves multiple indoor \acp{UE} randomly distributed within the coverage area. 
The \ac{2D} waveguide is deployed on a horizontal plane at height $z=h$ and accommodates $N$ reconfigurable \acp{PA}, all sharing a common \ac{RF} chain at the feed point. 
Unlike conventional \ac{PASS} or \acp{FPA}, the \ac{2D-PASS} offers substantially richer spatial \ac{DoFs} by enabling each \ac{PA} to slide along both the $x$- and $y$-axes, thereby achieving enhanced flexibility for beamforming and channel adaptation. 
Moreover, since the physical aperture of the dielectric waveguide typically extends over meter-scale dimensions, it naturally aligns with the spatial scale of typical indoor environments.
This intrinsic compatibility makes the \ac{2D-PASS} particularly attractive for indoor wireless applications, where compactness, configurability, and fine-grained spatial control are essential for efficient multiuser transmission.

Fig.~\ref{fig:system_model}(b) further illustrates the geometric structure of the \ac{2D} dielectric waveguide, which defines the feasible region $\mathcal{C}_r$ for \ac{PA} placement. 
Within this region, each \ac{PA} can move continuously along both the $x$- and $y$-axes, thereby offering substantially richer spatial \ac{DoFs} than conventional \ac{PASS} architectures based on line-shaped waveguides.

The feasible region $\mathcal{C}_r$ is mathematically defined as
\begin{equation}
	\mathcal{C}_r = \Big\{ [x,y,h]^{\rm T} \;\big|\; x \in [-\tfrac{D}{2},\tfrac{D}{2}],\; y \in [-\tfrac{D}{2},\tfrac{D}{2}] \Big\}.
	\label{Cr_def}
\end{equation}
The instantaneous configuration of all $N$ \acp{PA} is expressed as the stacked position vector
\begin{equation}
	{{\bm{\Psi }}^p} = \big[({\bm{\psi}}_1^p)^{\rm T}, ({\bm{\psi}}_2^p)^{\rm T}, \dots, ({\bm{\psi}}_N^p)^{\rm T}\big]^{\rm T},
	\label{Psi_def}
\end{equation}
where ${\bm{\psi}}_n^p=[x_n^p,y_n^p,h]^{\rm T}\in\mathcal{C}_r$ denotes the position of the $n$-th \ac{PA}. 
To mitigate excessive mutual coupling and mechanical interference, a minimum separation distance $D_0$ must be maintained between any two \acp{PA}, formulated as
\begin{equation}
	\|{\bm{\psi}}_i^p - {\bm{\psi}}_j^p\| \ge D_0, \quad \forall i \neq j.
	\label{D0_constraint}
\end{equation}

Without loss of generality, a \ac{LoS}-dominant propagation environment is considered, where the channel model accounts for spherical wavefront effects. 
The free-space path loss between the $n$-th \ac{PA} and the $k$-th \ac{UE} located at ${\bm{\psi}}_k^u = [x_k^u, y_k^u, 0]^{\mathrm{T}}$ is given by
\begin{equation}\label{path_loss}
	g({\bm{\psi}}_k^u ,{\bm{\psi}}_n^p) = \frac{\sqrt{\eta}}{\|{\bm{\psi}}_k^u - {\bm{\psi}}_n^p\|},
\end{equation}
where $\eta = \tfrac{\lambda_c^2}{16\pi^2}$ and $\lambda_c$ denotes the carrier wavelength. 
The corresponding propagation distance can be expressed as
\begin{equation}
	\|{\bm{\psi}}_k^u - {\bm{\psi}}_n^p\| 
	= \sqrt{(x_k^u - x_n^p)^2 + (y_k^u - y_n^p)^2 + h^2}.
	\label{distance}
\end{equation}

In addition to the distance-dependent attenuation, the overall channel coefficient also incorporates both the free-space propagation delay and the guided-wave delay within the dielectric waveguide. As illustrated in Fig.~\ref{fig:system_model}(b), the feed point is located at ${\bm{\psi}}_0 = [-D/2,\,0,\,0]^{\mathrm{T}}$.
Accordingly, the phase component of the channel coefficient can be represented as
\begin{equation}
	\begin{aligned}
	&h({\bm{\psi}}_k^u, {\bm{\psi}}_n^p)\\
	&= \exp\!\Big(-j\frac{2\pi \|{\bm{\psi}}_k^u - {\bm{\psi}}_n^p\|}{\lambda_c}\Big)
	\exp\!\Big(-j\frac{2\pi  \|{\bm{\psi}}_0 - {\bm{\psi}}_n^p\|}{\lambda_g}\Big),
	\label{channel_coeff}
		\end{aligned}
\end{equation}
where $\lambda_g = \tfrac{\lambda_c}{n_{\mathrm{eff}}}$ is the guided wavelength and $n_{\mathrm{eff}}$ denotes the effective refractive index of the dielectric waveguide. 
Specifically, the first exponential term in~\eqref{channel_coeff} represents the free-space phase shift, while the second term accounts for the in-waveguide propagation phase shift.

Furthermore, the channel is modeled as quasi-static block fading, remaining constant during each coherence block, while the reconfigurability of the \acp{PA} enables the \ac{BS} to adaptively reshape channel responses across different intervals.

Consequently, the equivalent channel coefficient between the $n$-th \ac{PA} and the $k$-th \ac{UE} can be expressed as
\begin{equation}\label{key}
	h_{k}^n = g({\bm{\psi}}_k^u, {\bm{\psi}}_n^p)\, h({\bm{\psi}}_k^u, {\bm{\psi}}_n^p),
\end{equation}
where $g(\cdot)$ and $h(\cdot)$ denote the large-scale attenuation and the composite phase response introduced by the free-space and guided-wave propagation, respectively.

\subsection{Signal Model}
Let $s$ denote the transmitted symbol drawn from an $M$-ary \ac{APM} constellation. We assume a equal power allocation across all \acp{PA} for simplicity. Therefore, the received signal at the $k$-th \ac{UE} is expressed as
\begin{equation}\label{receive_signal}
	\begin{aligned}
		y_k 
		&= \sqrt{\frac{P}{N}} \sum_{n=1}^{N} h_{k}^n s + n_k,\\
		&= \underbrace{\sqrt{\frac{P}{N}}
			\overbrace{\sum_{n=1}^{N} g(\bm{\psi}_k^u,\bm{\psi}_n^p)h(\bm{\psi}_k^u,\bm{\psi}_n^p)}^{\text{pinching beamforming}} s}_{\text{desired signal}}
		+ n_k,
	\end{aligned}
\end{equation}
where $ P $ denotes the transmit power and $n_k\sim\mathcal{CN}(0,\sigma^2)$ represents additive white Gaussian noise (AWGN). 
Accordingly, the received \ac{SNR} at the $k$-th \ac{UE} is given by
\begin{equation}
	\mathrm{SNR}_k({{\bm{\Psi }}^p})=\frac{P\big|\sum_{n=1}^N g({\bm{\psi}}^u_k,{\bm{\psi}}^p_n)h({\bm{\psi}}^u_k,{\bm{\psi}}^p_n)\big|^2}{N\sigma^2}.
	\label{snr_def}
\end{equation}
According to \eqref{receive_signal} and \eqref{snr_def}, it is evident that the received signal power is highly sensitive to the spatial configuration of the \acp{PA}. 
Therefore, by properly adjusting their locations, the system effectively realizes an analog beamforming effect~\cite{PASS_beamforming}, steering the beams toward desired \acp{UE} to enhance signal strength.

In summary, Fig.~\ref{fig:system_model} provides a unified representation of both the transmission scenario and the physical \ac{PA} deployment model. 
This joint view clarifies the interaction between the \ac{BS} and \acp{UE}, while highlighting the spatial flexibility of the \ac{2D-PASS}. 
The subsequent sections investigate continuous and discrete positioning schemes aiming at maximizing the minimum \ac{SNR} under constraint of the feasible region.

\section{Continuous 2D-PASS Design}\label{sec:continuous}
Unlike conventional \ac{PASS} and \ac{FPA} deployments, the proposed \ac{2D-PASS} permits every \ac{PA} to move continuously over the dielectric waveguide plane. This mobility unlocks additional spatial \ac{DoFs} for adaptive reshaping of multiuser channels. Therefore, optimizing the \acp{PA} in a truly \ac{2D} domain can significantly improve the worst received \ac{SNR} relative to traditional \ac{PASS} or \ac{FPA} architectures.

To tackle the ensuing nonconvex design, a \ac{PSO} strategy~\cite{MA4} is adopted for its derivative-free updates and strong global search capability. In the proposed \ac{PSO} framework, each candidate \ac{PA} layout is encoded as a particle that updates its position and velocity by combining \emph{cognitive} feedback (its personal best) with \emph{social} guidance (the swarm’s global best). This cooperation between self-exploration and collective learning drives the swarm toward high-quality configurations, making \ac{PSO} approach well suited to the coupled continuous-position optimization considered here.

\subsection{Problem Formulation}

Consider the downlink broadcast scenario illustrated in Fig.~\ref{fig:system_model}, 
where a \ac{BS} equipped with $N$ movable \acp{PA} simultaneously serves $K$ \acp{UE} positioned at known coordinates $\{{\bm{\psi}}_k^u\}_{k=1}^K$. 
The goal is to determine the continuous position vector ${{\bm{\Psi}}^p}$ of all \acp{PA} that maximizes the minimum received \ac{SNR} among all \acp{UE}, denoted by $\mu$. 
Accordingly, the continuous \ac{PA} positioning problem can be formulated as
\begin{subequations}\label{P1}
	\begin{align}
		(\mathrm{P} 1): & \max _{{\bm{\Psi}}^p, \mu}  \ \ \mu \\
		\text {s.t. } 
		& {\bm{\psi }}_n^p \in {{\cal C}_r}, \quad n=1,2, \ldots, N, \label{op1:sub1}\\
		& \|{\bm{\psi }}_i^p-{\bm{\psi }}_j^p\| \geq D_{0}, \quad 1 \le i \ne j \le N,\label{op1:sub2}\\
		& {\rm{SN}}{{\rm{R}}_k}({{\bm{\Psi }}^p}) \geq \mu, \quad k=1,2, \ldots, K,\label{op1:sub3}
	\end{align}
\end{subequations}
where constraint~\eqref{op1:sub1} confines each \ac{PA} within the feasible \ac{2D} region $\mathcal{C}_r$, 
while~\eqref{op1:sub2} enforces a minimum separation distance $D_0$ between any two \acp{PA} to mitigate electromagnetic coupling. 
Constraint~\eqref{op1:sub3} ensures that every \ac{UE} attains an \ac{SNR} not lower than the common threshold $\mu$. 

By jointly optimizing the continuous coordinates $\{{\bm{\psi}}_n^p\}_{n=1}^N$, 
the system can exploit high-resolution spatial diversity to enhance multiuser signal strength and overall performance. 
However, the strong nonlinear coupling among the continuous variables ${\bm{\psi}}_n^p$ makes problem (P1) a highly non-convex optimization task, 
as the \ac{SNR} constraints in~\eqref{op1:sub3} depend nonlinearly on the spatial positions of all \acp{PA}. 
To efficiently handle this challenging problem, the next subsection introduces a heuristic optimization framework based on \ac{PSO} approach.

\subsection{Proposed Algorithm}

To address the non-convex nature of problem (P1), the \ac{PSO} algorithm is adopted as an effective heuristic for optimizing the \ac{PA} positions in the continuous domain. 
\ac{PSO} framework has been extensively employed in non-differentiable and high-dimensional optimization tasks due to its strong global search ability, derivative-free formulation, and ease of implementation. 
Unlike gradient-based approaches, which often suffer from premature convergence to local optima, \ac{PSO} approach  maintains a favorable balance between exploration and exploitation by combining global collaboration and individual learning. 
Through this mechanism, the swarm collectively evolves toward high-quality solutions while preserving diversity, making \ac{PSO} strategy particularly well suited for the strongly coupled and highly non-linear position optimization problem investigated in this work.

\subsubsection*{1) Particle Encoding and Initialization}

A swarm consisting of $M$ particles is first generated to represent different candidate configurations of the $N$ \acp{PA}. 
Since all antennas are confined to the plane $z=h$, their $z$-coordinates remain constant and are therefore omitted, 
leaving only the $x$–$y$ coordinates within the feasible region $\mathcal{C}_r$ to be encoded. 
Accordingly, the $m$-th particle is characterized by its position vector
\begin{equation}
	{\bm{\Psi}}_{m}^{p,0}
	= \big[ ( {\bm{\psi}}_{m,1}^{p,0} )^{\rm T},\,
	( {\bm{\psi}}_{m,2}^{p,0} )^{\rm T},\,\ldots,\,
	( {\bm{\psi}}_{m,N}^{p,0} )^{\rm T} \big]^{\rm T},
\end{equation}
and its associated velocity vector
\begin{equation}
	{\bm{v}}_{m}^{p,0}
	= \big[ ( {\bm{v}}_{m,1}^{p,0} )^{\rm T},\,
	( {\bm{v}}_{m,2}^{p,0} )^{\rm T},\,\ldots,\,
	( {\bm{v}}_{m,N}^{p,0} )^{\rm T} \big]^{\rm T},
\end{equation}
where ${\bm{\psi}}_{m,n}^{p,0}=[x_{m,n}^{p,0},\,y_{m,n}^{p,0}]^{\rm T}$ and ${\bm{v}}_{m,n}^{p,0}\in\mathbb{R}^2$.

To improve convergence speed and the quality of the final solution, the initial particle positions must be carefully designed. 
Since the optimization objective is to maximize the minimum received \ac{SNR} across all \acp{UE}, and the path loss dominates the received signal power, 
it is desirable that each user initially has at least one nearby \ac{PA}. 
Therefore, a round-robin, user-centric initialization strategy is employed to distribute the \acp{PA} around users in a spatially balanced and diverse manner.

Let the horizontal coordinates of the $K$ users be ${\bm{\psi}}^u_k=[x_k^u,y_k^u,0]^{\rm T}$, where $k\in\{1,\ldots,K\}$.  
Each \ac{PA} index $n$ is cyclically assigned to users according to
\begin{equation}
	u(n) \triangleq 1 + \big((n-1) \bmod K\big),
\end{equation}
and the initial position of the $n$-th \ac{PA} in $ m $-th particle is randomly sampled within a disk of radius $r$ centered at the $u(n)$-th user:
\begin{align}
	\rho_{m,n} &= r\sqrt{\xi_{m,n}}, \qquad \xi_{m,n}\sim\mathcal{U}[0,1],\\
	\theta_{m,n} &\sim \mathcal{U}[0,2\pi),\\
	\widetilde{\bm{\psi}}_{m,n}^{p,0} 
	&= \begin{bmatrix} x_{u(n)}^u \\[2pt] y_{u(n)}^u \end{bmatrix}
	+ \rho_{m,n}\!\begin{bmatrix}\cos\theta_{m,n} \\ \sin\theta_{m,n}\end{bmatrix}.
\end{align}
To guarantee that all sampled positions lie within the feasible region $\mathcal{C}_r$ defined in~\eqref{Cr_def}, 
a projection operator ${\cal Q}(\cdot)$ is applied to clip any out-of-bound positions:
\begin{equation}
	{\bm{\psi}}_{m,n}^{p,0} = {\cal Q}\!\big(\widetilde{\bm{\psi}}_{m,n}^{p,0}\big),
\end{equation}
where ${\cal Q}(\cdot)$ operates element-wise as
\begin{equation}\label{key}
	{{\cal Q}({\bm{\Psi}}^p)} =
	\begin{cases}
		-\frac{D}{2}, & \text{if } {\bm{\Psi}}^p < -\frac{D}{2},\\[3pt]
		\frac{D}{2}, & \text{if } {\bm{\Psi}}^p> \frac{D}{2},\\[3pt]
		{\bm{\Psi}}^p, & \text{otherwise.}
	\end{cases}
\end{equation}

Moreover, if the minimum-distance constraint $\|{\bm{\psi}}_{m,i}^{p,0}-{\bm{\psi}}_{m,j}^{p,0}\|\ge D_0$ is violated, 
the conflicting positions are resampled until the configuration becomes feasible. 
This rejection–resampling process ensures that the initialized \acp{PA} are both user-centric and sufficiently dispersed within the feasible region, thereby avoiding coupling and overlap.

Finally, the initial velocities can be either set to zero or initialized with small random values, such as
\begin{equation}
	{\bm{v}}_{m,n}^{p,0} \sim \mathcal{U}\!\big([-\upsilon_{\max},\,\upsilon_{\max}]^2\big),
\end{equation}
which promotes early-stage exploration while maintaining numerical stability during the subsequent iterations.

\subsubsection*{2) Fitness Evaluation with Penalty Mechanism}

For each particle, the fitness value is defined as the minimum received \ac{SNR} among all \acp{UE}, 
while incorporating a penalty term to account for any violation of the minimum inter-\ac{PA} distance $D_0$ specified in~\eqref{D0_constraint}. 
Accordingly, the fitness of the $m$-th particle at iteration $t$ is expressed as
\begin{equation}\label{PSO_object}
	{\cal F}({\bm{\Psi}}_m^{p,t}) =
	\min_{k}\big\{\mathrm{SNR}_k({\bm{\Psi}}_m^{p,t})\big\}
	- \gamma\big|{\cal V}({\bm{\Psi}}_m^{p,t})\big|,
\end{equation}
where $\gamma$ denotes a sufficiently large penalty coefficient, and ${\cal V}(\cdot)$ represents the number of \ac{PA} pairs whose separation distance is smaller than $D_0$, i.e.,
\begin{equation}
	{\cal V}({\bm{\Psi}}_m^{p,t})
	\triangleq
	\big\{
	({\bm{\psi}}_i^p,{\bm{\psi}}_j^p)
	\,\big|\,
	\|{\bm{\psi}}_i^p - {\bm{\psi}}_j^p\| < D_0,
	\,1 \le i < j \le N
	\big\}.
\end{equation}
This penalty mechanism ensures that particles violating the spacing constraint receive a reduced fitness value, 
thereby discouraging infeasible configurations and guiding the swarm toward the feasible solution space during the optimization process.

\subsubsection*{3) Particle Update and Boundary Handling}

At each iteration, the fitness of all particles is evaluated, followed by the update of both the personal best position ${\bm{\Psi}}_{m,{\rm popt}}^{p,t}$ and the global best position ${\bm{\Psi}}_{{\rm gopt}}^{p,t}$. 
These best-known configurations are determined as
\begin{align}
	{\bm{\Psi}}_{m,{\rm popt}}^p 
	&= \arg\max\big\{F({\bm{\Psi}}_m^{p,t+1}),\,F({\bm{\Psi}}_{m,{\rm popt}}^p)\big\}, \label{eq:pbest}\\
	{\bm{\Psi}}_{{\rm gopt}}^p 
	&= \arg\max\big\{F({\bm{\Psi}}_m^{p,t+1}),\,F({\bm{\Psi}}_{{\rm gopt}}^p)\big\}. \label{eq:gbest}
\end{align}

Subsequently, each particle updates its velocity and position according to:
\begin{align}
	{\bm{v}}_m^{p,t+1} 
	&= \eta {\bm{v}}_m^{p,t} 
	+ \lambda_1 \omega_1 \big({\bm{\Psi}}_{m,{\rm popt}}^p - {\bm{\Psi}}_m^{p,t}\big) \notag\\
	&\quad + \lambda_2 \omega_2 \big({\bm{\Psi}}_{{\rm gopt}}^p - {\bm{\Psi}}_m^{p,t}\big),
	\label{update_v}\\[4pt]
	{\bm{\Psi}}_m^{p,t+1} 
	&= {\bm{\Psi}}_m^{p,t} + {\bm{v}}_m^{p,t+1},
	\label{update_Psi}
\end{align}
where $\eta$ denotes the inertia weight controlling the trade-off between global exploration and local exploitation, 
$\lambda_1$ and $\lambda_2$ are the acceleration coefficients associated with cognitive and social learning, 
and $\omega_1,\omega_2\!\sim\!\mathcal{U}[0,1]$ are independent random variables that introduce stochastic perturbations. 
These iterative updates guide the swarm toward promising regions of the search space while maintaining sufficient diversity to avoid premature convergence.

To further balance convergence speed and solution accuracy, a linearly decreasing inertia weight strategy is adopted, which can be expressed as
\begin{equation}
	\eta = \eta_{\max} - \frac{(\eta_{\max}-\eta_{\min})t}{T},
\end{equation}
where $\eta_{\max}$ and $\eta_{\min}$ represent the upper and lower bounds of $\eta$, respectively, and $T$ denotes the total number of iterations. 
This adaptive scheduling enables broad global exploration during the early stages and gradual refinement of high-quality solutions as the search progresses.

After each position update, a boundary-handling mechanism ensures that all \acp{PA} remain within the feasible deployment region. 
To mitigate excessive path loss and enhance the achievable minimum SNR, the updated coordinates are restricted within a rectangular region centered around the \acp{UE}. 
Let $x_{\min}$, $x_{\max}$, $y_{\min}$, and $y_{\max}$ represent the minimum and maximum spatial coordinates of the \acp{UE} along the $x$- and $y$-axes, respectively, which are given by
\begin{equation}
	\begin{aligned}
		x_{\min} &= \min_k x_k^u,\quad
		x_{\max} = \max_k x_k^u,\\
		y_{\min} &= \min_k y_k^u,\quad
		y_{\max} = \max_k y_k^u.
	\end{aligned}
\end{equation}

Accordingly, the feasible region is defined as
\begin{equation}
	\mathcal{C}_u = \big\{ (x,y)\mid x\!\in\![x_{\min},x_{\max}],\; y\!\in\![y_{\min},y_{\max}] \big\}.
\end{equation}

Furthermore, after updating the particle positions, each element of ${\bm{\Psi}}_{m}^{p,t+1}$ is projected into the feasible region $\mathcal{C}_u$ through a coordinate-wise projection operator ${\cal G}_u(\cdot)$. 
This operator independently clips the $x$- and $y$-coordinates of each \ac{PA} to their respective boundary limits, ensuring that all updated positions remain within the valid operational area. 
Specifically, ${\cal G}_u^x(\cdot)$ and ${\cal G}_u^y(\cdot)$ constrain the $x$-axis and $y$-axis components of ${\bm{\Psi}}^p$ to the ranges $[x_{\min},x_{\max}]$ and $[y_{\min},y_{\max}]$, respectively, which can be represented as
\begin{equation}\label{clip1}
	\begin{aligned}
		[{\cal G}_u^x({\bm{\Psi}}^p)]_i &=
		\begin{cases}
			x_{\min}, & \text{if } [{\bm{\Psi}}^p_x]_i < x_{\min},\\[3pt]
			x_{\max}, & \text{if } [{\bm{\Psi}}^p_x]_i > x_{\max},\\[3pt]
			[{\bm{\Psi}}^p_x]_i, & \text{otherwise,}
		\end{cases}\\[6pt]
		\end{aligned}
\end{equation}
and
\begin{equation}\label{clip2}
	\begin{aligned}
		[{\cal G}_u^y({\bm{\Psi}}^p)]_i &=
		\begin{cases}
			y_{\min}, & \text{if } [{\bm{\Psi}}^p_y]_i < y_{\min},\\[3pt]
			y_{\max}, & \text{if } [{\bm{\Psi}}^p_y]_i > y_{\max},\\[3pt]
			[{\bm{\Psi}}^p_y]_i, & \text{otherwise.}
		\end{cases}
	\end{aligned}
\end{equation}

This coordinate-wise projection mechanism guarantees that all updated antenna positions remain confined within the \ac{UE}-centric feasible region $\mathcal{C}_u$, thereby preventing out-of-bound movements, mitigating excessive propagation loss, and improving both the robustness and stability of the overall optimization process.

\subsubsection*{4) Termination and Solution Selection}

The iterative process proceeds until the maximum number of iterations $T$ is reached or the improvement in the global best fitness value falls below a predefined threshold. Upon termination, the global best position is adopted as the optimized configuration of the \acp{PA}, expressed as
\begin{equation}
	{\bm{\Psi}}^p = {\bm{\Psi}}_{{\rm gopt}}^p.
\end{equation}
The corresponding minimum received \ac{SNR} is subsequently calculated as $\mu = \min_{k} \mathrm{SNR}_k({\bm{\Psi}}^p)$.

Through iterative collaboration among all particles—each refining its own best position based on individual experience and the collective knowledge of the swarm—the global best solution gradually converges to a stable value. This cooperative learning process ultimately yields a high-quality continuous positioning strategy for the proposed \ac{2D-PASS}.

\subsection{Overall Framework and Complexity Analysis}
\begin{figure}[t]
	\renewcommand{\algorithmicrequire}{\textbf{Input:}}
	\renewcommand{\algorithmicensure}{\textbf{Output:}}
	\begin{algorithm}[H]
		\caption{The proposed \ac{PSO}-Based Continuous Positioning Algorithm for \ac{2D-PASS}}
		\begin{algorithmic}[1]
			\REQUIRE $N$, $K$, $\{{\bm{\psi}}_k^u\}_{k=1}^K$, $D$, $D_0$, $\gamma$, $\eta$, $\lambda_1$, $\lambda_2$, $\omega_1$, $\omega_2$.
			\ENSURE ${\bm{\Psi}}^p$, $\mu$.
			\STATE Randomly initialize the positions ${\bm{\Psi}}_m^{p,0}$ and velocities ${\bm{v}}_m^{p,0}$ of all $M$ particles.
			\STATE Evaluate the initial fitness value of each particle using~\eqref{PSO_object}.
			\FOR{$t = 1,2,\dots,T$}
			\FOR{$m = 1,2,\dots,M$}
			\STATE Update ${\bm{v}}_m^{p,t+1}$ and ${\bm{\Psi}}_m^{p,t+1}$ via~\eqref{update_v},~\eqref{update_Psi},~\eqref{clip1} and~\eqref{clip2}.
			\STATE Compute the new fitness value $	{\cal F}({\bm{\Psi}}_m^{p,t+1})$ using~\eqref{PSO_object}.
			\STATE Update the personal best position:
			\[
			{\bm{\Psi}}_{m,{\rm popt}}^p = \arg\max\big\{	{\cal F}({\bm{\Psi}}_m^{p,t+1}),\,	{\cal F}({\bm{\Psi}}_{m,{\rm popt}}^p)\big\}.
			\]
			\STATE Update the global best position:
			\[
			{\bm{\Psi}}_{{\rm gopt}}^p = \arg\max\big\{	{\cal F}({\bm{\Psi}}_m^{p,t+1}),\,	{\cal F}({\bm{\Psi}}_{{\rm gopt}}^p)\big\}.
			\]
			\ENDFOR
			\ENDFOR
			\STATE Set ${\bm{\Psi}}^p = {\bm{\Psi}}_{{\rm gopt}}^p$.
			\STATE Compute $\mu = \min_{k} \mathrm{SNR}_k({\bm{\Psi}}^p)$.
			\RETURN ${\bm{\Psi}}^p$, $\mu$.
		\end{algorithmic}
		\label{alg:alg1}
	\end{algorithm}
\end{figure}

The overall procedure of the proposed \ac{PSO}-based algorithm for solving problem (P1) is summarized in \textbf{Algorithm~\ref{alg:alg1}}. 
Initially, all particles are randomly generated with different positions and velocities, and their fitness values are evaluated based on the objective function defined in~\eqref{PSO_object}. 
During each iteration, every particle updates its position and velocity according to~\eqref{update_v}, \eqref{update_Psi}, ~\eqref{clip1} and~\eqref{clip2}, which incorporates both its own historical best position and the global best position discovered by the swarm. 
After each update, the corresponding fitness value is recalculated, and the best positions are refined accordingly. 
Through this iterative process, the swarm continuously evolves toward more promising regions of the search space, leading to a monotonic improvement in the global best fitness value until convergence. 
Since the stochastic initialization of particles may affect the final solution, the optimization can be executed multiple times with different random seeds, and the best result among all runs is selected as the final output. 
This procedure ensures obtaining a high-quality suboptimal positioning solution  ${\bm{\Psi}}^p$, while achieving a satisfactory minimum \ac{SNR} value $\mu$.

Subsequently, the computational complexity of the proposed continuous-positioning algorithm is analyzed as follows. 
In each iteration, the velocity and position updates for $N$ pinching \acp{PA} require $\mathcal{O}(N)$ floating-point operations per particle, whereas the pairwise distance verification for the minimum inter-\ac{PA} spacing introduces an additional $\mathcal{O}(N^2)$ cost. 
Moreover, evaluating the received \ac{SNR} for $K$ \acp{UE} involves summing $N$ complex coefficients per \ac{UE}, resulting in $\mathcal{O}(KN)$ operations. 
Hence, the overall computational cost per particle per iteration is $\mathcal{O}(N^2 + KN)$. 
Given that the swarm consists of $M$ particles and the algorithm runs for $T$ iterations, the total computational complexity can be expressed as
\begin{equation}
	\mathcal{O}\!\left(MT(N^2 + KN)\right),
	\label{eq:complexity}
\end{equation}
which scales linearly with the swarm size $M$ and iteration count $T$, and quadratically with the number of \acp{PA} $N$. 
This confirms that the proposed \ac{PSO}-based optimization framework achieves a favorable balance between computational efficiency and solution accuracy, making it well-suited for practical real-time implementation in \ac{2D-PASS} systems of moderate scale.

\section{Discrete 2D-PASS Design}\label{sec:discrete}
\begin{figure}[t]
	\centering
	\includegraphics[width=0.7\linewidth]{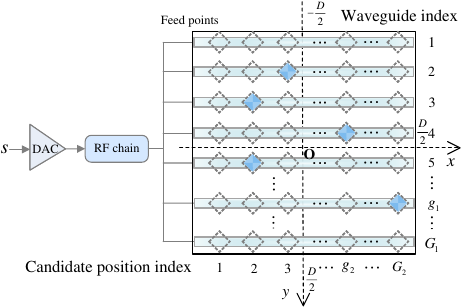}
	\caption{Illustration of the \ac{2D-PASS} with discrete candidate positions.}
	\label{discrete}
	
\end{figure}
Although the continuous positioning scheme developed in Section~III can provide a considerable performance, its practical deployment is severely constrained by hardware limitations. In real \ac{2D-PASS} implementations, the movement of each \ac{PA} is typically confined to a finite set of mechanically predefined pinching slots or electrically switchable contact points on the dielectric waveguide. These hardware constraints make arbitrary continuous repositioning impractical, while the associated control overhead increases prohibitively with the number of \acp{PA}.

These considerations motivate the design of a discrete positioning scheme that achieves improved performance while remaining compatible with practical hardware implementations. To this end, we discretize the feasible \ac{2D} region $\mathcal{C}_r$ into a finite grid of candidate positions and reformulate the \ac{PA} positioning task as a combinatorial optimization problem that selects the optimal subset of positions. This discrete modeling approach naturally transforms the original continuous geometric optimization into a binary decision-making task, where each grid point is either activated to host a \ac{PA} or left unoccupied. However, the resulting problem involves binary selection variables and pairwise separation constraints, which together render it a large-scale and highly coupled \ac{MINLP} problem.

In the following, we first present the problem formulation of this discrete positioning scheme and then develop an efficient \ac{MILP}-based algorithm to solve it.

%

\subsection{Problem Formulation}
While modeling the \ac{PA} positions as continuous variables offers a convenient optimization framework, it neglects several crucial constraints inherent in practical implementations. In real systems, most existing \ac{PASS} prototypes deploy \acp{PA} along line-shaped dielectric waveguides. Therefore, a more realistic implementation of the proposed \ac{2D-PASS} is to arrange multiple parallel dielectric waveguides to approximate a \ac{2D} plane, as depicted in Fig.~\ref{discrete}.  This architecture inherently quantizes the antenna positions along the dimension perpendicular to the waveguides. Moreover, even along the longitudinal axis of each waveguide, the achievable positions are effectively discretized due to the finite actuation precision of the electromechanical control system. Consequently, a discrete positioning model is more representative of the physical reality and thus of greater practical value.

In light of these considerations, we assume that $G_1$ dielectric waveguides are uniformly deployed along the $y$-axis with an equal quantization step  of $\Delta_1 = D/(G_1-1)$ between adjacent waveguides. Each waveguide supports $G_2$ uniformly spaced candidate positions along its length with quantization step  of $\Delta_2 = D/(G_2-1)$. By indexing these positions in a linear fashion, the $g_2$-th ($g_2 \in \{1,2,\dots,G_2\}$) candidate position on the $g_1$-th ($g_1 \in \{1,2,\dots,G_1\}$) waveguide is mapped to a global index as
\begin{equation}
	g = (g_1-1)G_2+g_2.
\end{equation}
Accordingly, the \ac{3D} coordinate of the $g$-th candidate position ($g\in\{1,2,\dots,G\}$, $G = G_1\times G_2$) is given by
\begin{equation}\label{pg_definition}
	\mathbf{p}_{g} = \left[-\frac{D}{2} + \frac{(g_2-1)D}{G_2-1}, -\frac{D}{2} + \frac{(g_1-1)D}{G_1-1},  h \right]^{\rm T},
\end{equation}
with $\mathcal{P} = \{\mathbf{p}_1, \mathbf{p}_2, \dots, \mathbf{p}_G\} \subseteq \mathcal{C}_r$ denotes the set of all $G$ candidate positions. Correspondingly, the full channel vector for the $k$-th user is expressed as
\begin{equation}
	\mathbf{h}_k = [h_{k,1}, h_{k,2}, \dots, h_{k,G}]^{\rm T},
\end{equation}
where $h_{k,g} = g({\bm{\psi}}_k^u,\mathbf{p}_{g})\,h({\bm{\psi}}_k^u,\mathbf{p}_{g})$ denotes the channel coefficient between the $g$-th candidate position and the $k$-th user. 
As illustrated in Fig.~\ref{discrete}, the discrete configuration differs from the continuous case in that $G_1$ dielectric waveguides are deployed, 
where the feed point of the $g_1$-th waveguide is located at the coordinates ${[-\frac{D}{2},\,-\frac{D}{2} + \frac{(g_1-1)D}{G_1-1},\,h]^{\mathrm{T}}}$.
Accordingly, the phase component of the channel coefficient in the discrete case can be expressed as
\begin{equation}
	h({\bm{\psi}}_k^u, \mathbf{p}_{g})
	= \exp\!\Big(-j\frac{2\pi \|{\bm{\psi}}_k^u - \mathbf{p}_{g}\|}{\lambda_c}\Big)
	\exp\!\Big(-j\frac{2\pi (x_{g} + \frac{D}{2})}{\lambda_g}\Big),
	\label{channel_coeff_discrete}
\end{equation}
where $x_{g}$ denotes the $x$-coordinate component of ${\bf{p}}_{g}$.


Under this discrete position model, our task is to select $N$ positions from the $G$ available candidates to deploy the \acp{PA}. This process can be described by binary selection vector $\mathbf{b} = [b_1, b_2, \dots, b_G]^{\rm T}\in \{0,1\}^{G \times 1}$, whose elements are given by
\begin{equation}
	b_g = 
	\begin{cases}
		1, & \text{if the candidate position } \mathbf{p}_{g} \text{ is selected,} \\
		0, & \text{otherwise.}
	\end{cases}
\end{equation}
To ensure that exactly $N$ positions are selected, the selection vector must satisfy the constraint $\sum_{g=1}^{G} b_g = N$. Furthermore, to guarantee the minimum separation distance $D_0$ between any two selected positions, the set of all conflicting position pairs is first defined as
\begin{equation}\label{conflictSet}
	\mathcal{S} = \left\{ (i, j) \mid 1 \le i < j \le G, \ |\mathbf{p}_i - \mathbf{p}_j| < D_0 \right\}.
\end{equation}
By prohibiting the simultaneous selection of any two conflicting positions, the minimum distance constraint~\eqref{op1:sub2} translates into a linear constraint on the selection variables as
\begin{equation}
	b_i + b_j \le 1, \quad \forall (i, j) \in \mathcal{S}.
\end{equation}

Given the position selection vector $\mathbf{b}$, the received signal in~\eqref{receive_signal} can be represented as
\begin{equation}
	\label{receive_signal_discrete}
	y_k = \frac{1}{N} \mathbf{h}_k^{\rm T}\mathbf{b}s  + n_k.
\end{equation}
Similar to~$(\mathrm{P}1)$, the objective here is to determine the optimal \ac{PA} placement that maximizes the minimum \ac{SNR} among all $K$ users. Consequently, problem~$(\mathrm{P}1)$ can be reformulated as the following discrete position selection problem:
\begin{subequations}\label{P2}
	\begin{align}
		(\mathrm{P} 2): & \max _{\mathbf{b}, \mu} \ \ \mu \\
		\text { s.t. } & b_g \in \{0,1\}, \quad g=1,2, \ldots, G, \label{op2:sub1} \\
		& \sum_{g=1}^{G} b_g = N, \label{op2:sub2} \\
		& b_i + b_j \le 1, \quad \forall (i, j) \in \mathcal{S}, \label{op2:sub3} \\
		& \frac{{| \mathbf{h}_k^{\rm T}\mathbf{b} |}^2}{N^2 \sigma^2} \ge \mu, \quad k=1,2, \ldots, K. \label{op2:sub4}
	\end{align}
\end{subequations}
However, problem $(\mathrm{P} 2)$ is a large-scale \ac{MINLP}, which is NP-hard due to the binary decision variables and the quadratic \ac{SNR} constraints. Its combinatorial search space and strong variable coupling make it non-convex and highly intractable to solve directly. In the next subsection, we reformulate $(\mathrm{P} 2)$ into an equivalent \ac{MILP} problem that can be solved to global optimality using off-the-shelf solvers.

\subsection{Proposed Algorithm}
To enable efficient computation, problem~$(\mathrm{P}2)$ is transformed into a more tractable \ac{MILP} formulation. The key idea is to linearize the quadratic term $|\mathbf{h}_k^{\rm T}\mathbf{b}|^2$ in the \ac{SNR} constraint~\eqref{op2:sub4}. Specifically, leveraging the property that $\mathbf{b}$ is binary (i.e., $b_i^2 = b_i$), the received signal power for the $k$-th user can be expanded as
\begin{align}
	|\mathbf{h}_k^{\rm T}\mathbf{b}|^2 &=\mathbf{b}^T\mathbf{h}_k^*\mathbf{h}_k^{\rm T}\mathbf{b}= \sum_{i=1}^{G} \sum_{j=1}^{G} h_{k,i}^* h_{k,j} b_i b_j \nonumber \\
	&= \sum_{i=1}^{G} |h_{k,i}|^2 b_i + \sum_{1 \le i < j \le G} 2\text{Re}\{h_{k,i}^* h_{k,j}\} b_i b_j. \label{eq:expanded_snr}
\end{align}
To handle the quadratic binary products $b_i b_j$ in~\eqref{eq:expanded_snr}, auxiliary binary variables $z_{i,j}\in \{0,1\}$ are introduced to enforce $z_{i,j} = b_i b_j$ for all $1 \le i < j \le G$. Specifically, this equivalence can be established by the McCormick relaxation constraints~\cite{mccormick} as
\begin{subequations}\label{eq:mccormick}
	\begin{align}
		& z_{i,j} \le b_i, \quad \forall 1 \le i < j \le G, \label{eq:mc1}\\
		& z_{i,j} \le b_j, \quad \forall 1 \le i < j \le G, \label{eq:mc2}\\
		& z_{i,j} \ge b_i + b_j - 1, \quad \forall 1 \le i < j \le G. \label{eq:mc3}
	\end{align}
\end{subequations}
These constraints guarantee that $z_{i,j}$ equals 1 if and only if both $b_i$ and $b_j$ are 1, and 0 otherwise. Moreover, substituting $z_{i,j}$ for $b_i b_j$ in~\eqref{eq:expanded_snr} yields a linearized expression of the received signal power, which can be given by
\begin{equation}
	\Gamma_k(\mathbf{b}, \mathbf{z}) \triangleq \sum_{i=1}^{G} |h_{k,i}|^2 b_i + \sum_{1 \le i < j \le G} 2\text{Re}\{h_{k,i}^* h_{k,j}\} z_{i,j}. \label{eq:linear_snr_func}
\end{equation}

By incorporating the auxiliary variables $z_{i,j}$ and the corresponding constraints, problem~$(\mathrm{P}2)$ can be reformulated as the following \ac{MILP}:
\begin{subequations}\label{P3}
	\begin{align}
		(\mathrm{P} 3): & \max _{\mathbf{b}, \mathbf{z}, \mu} \ \ \mu \\
		\text { s.t. } & z_{i,j} \in \{0,1\}, \quad \forall 1 \le i < j \le G, \\
		& \Gamma_k(\mathbf{b}, \mathbf{z}) \ge N^2 \sigma^2 \mu, \quad k=1,2, \ldots, K,\\
		& \eqref{op2:sub1}, \eqref{op2:sub2}, \eqref{op2:sub3}, \eqref{eq:mc1}, \eqref{eq:mc2}, \eqref{eq:mc3}.\nonumber
	\end{align}
\end{subequations}
Particularly, problem~$(\mathrm{P}3)$ is an \ac{MILP} that can be solved to global optimality using state-of-the-art optimization solvers such as Gurobi~\cite{gurobi}. Despite its worst-case exponential complexity, this class of problems has been extensively studied and can be efficiently solved via the branch-and-cut framework~\cite{gurobi}. Notably, when the number of candidate positions is sufficiently large (i.e., under fine-grained spatial quantization), the solution to~$(\mathrm{P}3)$ serves as a tight approximation to the global optimum of the original continuous problem~$(\mathrm{P}1)$. This makes it a valuable benchmark for evaluating the performance of other suboptimal algorithms and for assessing the theoretical limits of the proposed \ac{2D-PASS} system.

\subsection{Overall Framework and Complexity Analysis}
\begin{figure}[t]
	\renewcommand{\algorithmicrequire}{\textbf{Input:}}
	\renewcommand{\algorithmicensure}{\textbf{Output:}}
	\begin{algorithm}[H]
		\caption{The proposed \ac{MILP}-Based Discrete Positioning Algorithm for \ac{2D-PASS}}
		\begin{algorithmic}[1]
			\REQUIRE $N$, $K$, $\{\bm{\psi}_k^u\}_{k=1}^K$, $D$, $D_0$, $G_1$, $G_2$, $\sigma^2$.
			\ENSURE $\bm{b}^\star$, $\mu$.
			\STATE Initialize the candidate set $\mathcal{P}=\{{{\bf{p}}_1},\dots,{{\bf{p}}_G}\}$ via~\eqref{pg_definition}.
			\STATE Evaluate the channel coefficients $h_{k,g}=g(\bm{\psi}_k^u,{{\bf{p}}_g})h(\bm{\psi}_k^u,{{\bf{p}}_g})$ and assemble the channel vectors ${\bf h}_k=[h_{k,1},\dots,h_{k,G}]^{\rm T}$ for all $k\in\{1,\dots,K\}$.
			\STATE Obtain the conflict set $\mathcal{S}$ as~\eqref{conflictSet}.
			\STATE Define the binary variables $\{b_g\}$ and $\{z_{i,j}\}$, and formulate the \ac{MILP} optimization problem~(P3) according to~\eqref{P3}.
			\STATE Solve problem~(P3) using an MILP solver to determine the optimal selection vector $\bm{b}^\star$ and the resulting \ac{SNR} $\mu$.
			\RETURN $\bm{b}^\star$, $\mu$.
		\end{algorithmic}
		\label{alg:alg2}
	\end{algorithm}
\end{figure}

The proposed algorithm employs an \ac{MILP}-based framework to solve problem~(P2) with guaranteed optimality, as summarized in \textbf{Algorithm~\ref{alg:alg2}}. Specifically, a \ac{2D} grid comprising $G = G_1 \times G_2$ candidate positions is first constructed over the dielectric waveguide. The channel coefficients between each candidate position and every user are then computed to form the complete channel matrix. Based on these coefficients, the set of conflicting candidate pairs is identified according to the minimum inter-\ac{PA} distance constraint. Subsequently, auxiliary variables $\{z_{i,j}\}$ are introduced to transform the original \ac{MINLP} problem~(P2) into the equivalent \ac{MILP} problem~(P3). Finally, problem~(P3) is solved by a commercial \ac{MILP} solver (e.g., Gurobi), which yields the optimal discrete \ac{PA} placement ${\bm b}^\star$ and the corresponding minimum \ac{SNR} $\mu$.

In terms of computational complexity, problem (P2) is an NP-hard combinatorial optimization problem. Finding the optimal solution through exhaustive search involves enumerating a total of $\binom{G}{N}$ possible \ac{PA} placements. For each placement, one must verify its feasibility against the distance constraint~\eqref{op2:sub3} and then compute the objective value $\mu$ according to~\eqref{op2:sub4}. This mainly requires calculating the SNR for all $K$ users, where each SNR computation takes $\mathcal{O}(G)$ time. The cost of evaluating a single placement is therefore $\mathcal{O}(KG)$. Consequently, the overall complexity of an exhaustive search is given by $\mathcal{O}\left(\binom{G}{N}KG\right)$, which becomes computationally intractable even for moderate values of $G$ and $N$.

By reformulating (P2) into the equivalent \ac{MILP} problem (P3), we can utilize advanced commercial solvers. While the worst-case complexity for solving such an \ac{MILP} remains exponential, modern solvers could handle the problem efficiently using the branch-and-cut framework. Specifically, this framework systematically explores the search tree by solving linear programming (LP) relaxations at each node and introducing cutting planes to significantly prune the search space. A precise complexity analysis of this process is generally intractable, as the effectiveness of the pruning process is sensitive to the specific numerical structure of the problem instance. In practice, however, the branch-and-cut framework has proven to be exceptionally efficient for the MILP problems. Although this \ac{MILP}-based approach incurs a higher computational cost than heuristic algorithms, its ability to find a globally optimal solution makes it an essential benchmark for assessing the theoretical performance limits of the proposed \ac{2D-PASS} system.

\section{Simulation Results}\label{sec:simulation}
\begin{table}[!t]
	\centering
	\small
	\renewcommand{\arraystretch}{1.3}
	\caption{SIMULATION PARAMETERS OF 2D-PASS}
	\begin{tabular}{>{\centering\arraybackslash}p{4.8cm} >{\centering\arraybackslash}p{6.0cm}} 
		\toprule
		\textbf{Notation} & \textbf{Values} \\
		\midrule
		Number of PAs & $N=\{4,8,12,16,20\}$ \\
		Number of UEs & $K=\{4,5,6,7,8\}$ \\
		Transmit power & $P=\{10,15,20,25,30\}\,\mathrm{dBm}$ \\
		Noise power & $P_n=-80\,\mathrm{dBm}$ \\
		Side length & $D=\{20,30,40,50\}\,\mathrm{m}$ \\
		Carrier frequency & $f_c=28\,\mathrm{GHz}$ \\
		Height of antennas & $h=3\,\mathrm{m}$ \\
		Minimum PA distance & $D_0=\lambda_c/2$ \\
		Effective refractive index & $n_{\mathrm{eff}}=1.4$ \\
		Coordinate of the feed point & $\boldsymbol{\phi}_0=[-D/2,0,h]^{\rm{T}}$ \\
		\bottomrule
	\end{tabular}
	\label{tab:parameters}
\end{table}

\begin{table}[!t]
	\centering
	\small
	\renewcommand{\arraystretch}{1.3}
	\caption{SIMULATION PARAMETERS OF PSO}
	\begin{tabular}{>{\centering\arraybackslash}p{3.8cm} >{\centering\arraybackslash}p{4.0cm}} 
		\toprule
		\textbf{Notation} & \textbf{Values} \\
		\midrule
		Acceleration coefficients & $\lambda_1=\lambda_2=1.5$ \\
		Inertia weight & $\eta_{\max}=0.9$, $\eta_{\min}=0.4$ \\
		Penalty factor & $\gamma=30$ \\
		Number of particles & $M=\{500,1000\}$ \\
		Number of iterations & $T=200$ \\
		Radius of initialization & $r=2\,\mathrm{m}$ \\
		\bottomrule
	\end{tabular}
	\label{tab:pso_parameters}
\end{table}

In this section, simulation results are provided to validate the effectiveness of the proposed \ac{2D-PASS} schemes, encompassing both the continuous and discrete positioning designs. 
The system performance is evaluated in terms of the minimum received \ac{SNR} among all \acp{UE}, which is defined as ${\rm SNR} = \min_k {\rm SNR}_k$. 
The $K$ \acp{UE} are assumed to be randomly and uniformly distributed within a square area of side length $D$, centered at $[0,0,0]^{\rm T}$, as illustrated in Fig.~\ref{fig:system_model}(a). Therefore, the results are obtained by averaging over 500 independent random realizations of \ac{UE} distributions.
The detailed simulation parameters and algorithmic configurations are summarized in Table~\ref{tab:parameters} and Table~\ref{tab:pso_parameters}, respectively. 
For the discrete design, an identical quantization step is adopted along both axes, i.e., $\Delta_1 = \Delta_2 = \Delta$, to simplify the simulation process. 
The specific parameter values are explicitly indicated in the corresponding simulation scenarios.

For performance comparison, the following benchmark schemes are considered in the simulations:
\begin{itemize}
	\item \textbf{Proposed continuous \ac{2D-PASS}}: The continuous-position design introduced in Section~\ref{sec:continuous}.
	
	\item \textbf{Proposed discrete \ac{2D-PASS}}: The discrete-position design presented in Section~\ref{sec:discrete}.
	
	\item \textbf{Conventional PASS (Con-PASS)}: In this benchmark, the configuration of the \acp{PA} and the waveguide follows that of the system presented in~\cite{PA_DL_RateMax}. Specifically, the \acp{PA} are deployed along an line-shaped dielectric waveguide positioned at $y = 0$ and height $h = 3~\mathrm{m}$.  For a fair comparison, the \acp{PA} are continuously adjustable along the waveguide, while the numbers of \acp{PA} and \ac{RF} chains are identical to those adopted in the proposed \ac{2D-PASS}. In addition, the \ac{PA} positioning design follows the same \ac{PSO}-based algorithm introduced in Section~\ref{sec:continuous}.
	
\item \textbf{Fixed-position antenna (FPA)}: 
In this benchmark, a \ac{ULA} is deployed along the $x$-axis at a height of $h = 3~\mathrm{m}$, with its center positioned at $[0,0,d]^{\mathrm{T}}$, consistent with the antenna configuration described in~\cite{ULA}. 
The antenna elements are uniformly spaced by a distance of $D_0$. 
For a fair comparison, the total number of antennas is set identical to that in the proposed \ac{2D-PASS} configuration, with an equal number of \ac{RF} chains assigned. 
Moreover, the same precoding design as that in~\cite{MA_mutibeam} is employed.

\end{itemize}

\begin{figure}[t]
	\centering
	\includegraphics[width=0.65\linewidth]{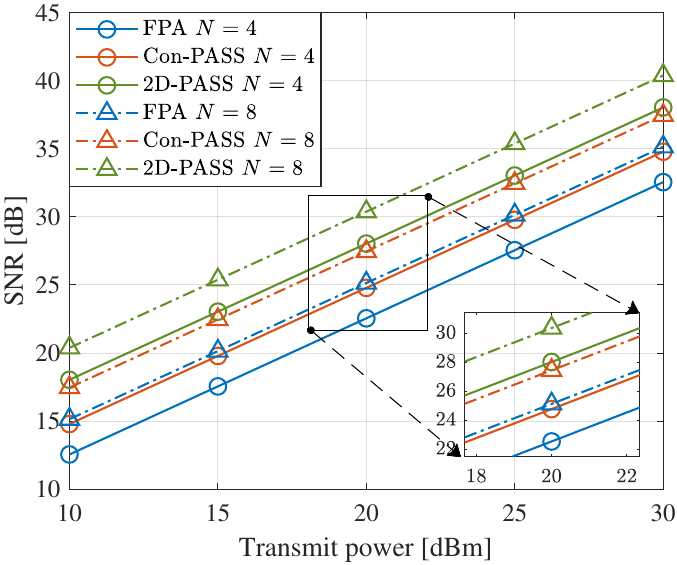}
	\caption{Minimum SNR versus transmit power for different schemes with $K=4$, $D=20$ m, and $N=\{4,8\}$.}
	\label{compare_P}
\end{figure}

As illustrated in Fig.~\ref{compare_P}, the minimum received \ac{SNR} among all \acp{UE} is plotted versus the transmit power for different schemes, with $K = 4$, $D = 20$~m, and $N = \{4, 8\}$. 
The Con-PASS and proposed \ac{2D-PASS} adopt the continuous-position design. 
It is evident that the performance of all schemes improves with increasing $N$, since a larger number of \acp{PA} provides additional spatial \ac{DoFs} for beamforming optimization. 
In particular, for a fixed antenna configuration, the proposed \ac{2D-PASS} consistently outperforms both the conventional \ac{PASS} and the \ac{FPA}. 
For instance, when $N = 4$ and $P = 20$~dBm, the \ac{2D-PASS} achieves approximately $3$~dB and $5$~dB \ac{SNR} gains over the conventional \ac{PASS} and \ac{FPA}, respectively. 
This performance advantage originates from the ability of the \ac{2D-PASS} to effectively mitigate path loss through flexible \ac{2D} spatial reconfiguration, thereby validating the superiority and practical effectiveness of the proposed architecture.

\begin{figure}[t]
	\centering
	\includegraphics[width=0.65\linewidth]{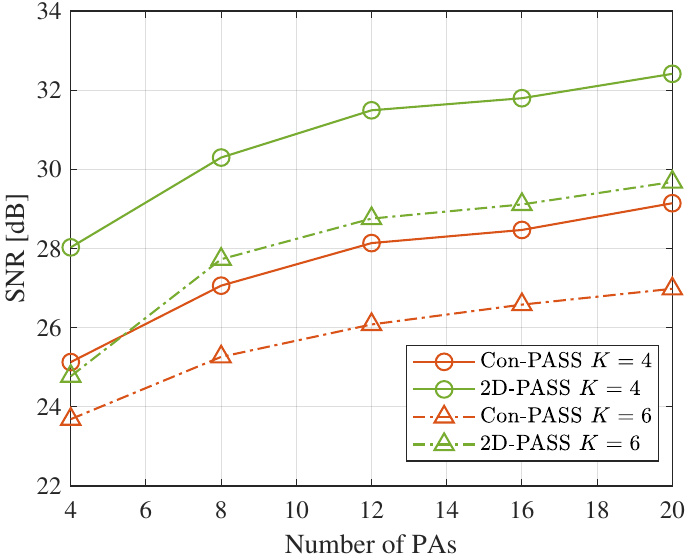}
	\caption{Minimum SNR versus the number of \acp{PA} for different schemes with $P=20$ dBm, $D=20$ m, and $K=\{4,6\}$.}
	\label{compare_N}
\end{figure}

Fig.~\ref{compare_N} presents the variation of the minimum received \ac{SNR} with respect to the number of \acp{PA} under different schemes, where $P = 20$~dBm, $D = 20$~m, and $K = \{4,6\}$. 
The Con-PASS and proposed \ac{2D-PASS} adopt the continuous-position design. 
It is observed that as the number of \acp{UE} increases, the overall \ac{SNR} performance declines, which can be attributed to the reduced transmit power allocated to each \ac{UE}. 
Meanwhile, for a fixed user number, the proposed \ac{2D-PASS} consistently achieves superior performance compared to the conventional \ac{PASS}. 
For example, when $N = 8$ and $K = 6$, the \ac{2D-PASS} attains an approximately $2.5$~dB improvement over its conventional counterpart. 
This gain arises from the richer spatial \ac{DoFs} enabled by the \ac{2D} mobility of the \acp{PA}, which enhances beam steering flexibility and mitigates path loss. 
Hence, the proposed \ac{2D-PASS} proves particularly effective in accommodating multi-user transmission scenarios.

\begin{figure}[t]
	\centering
	\includegraphics[width=0.65\linewidth]{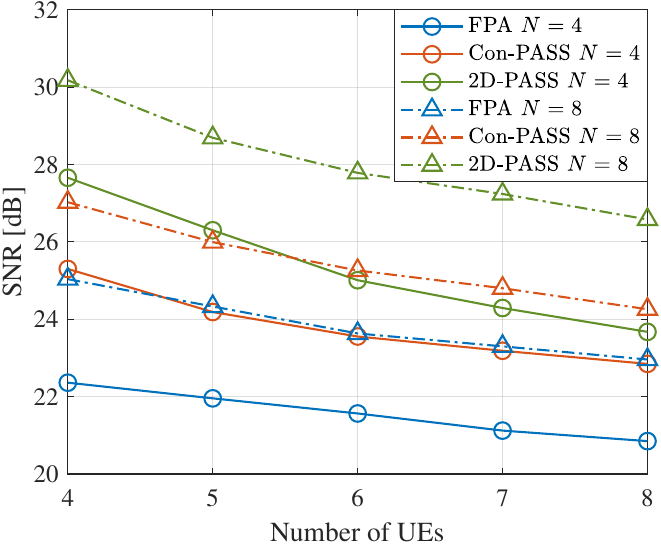}
	\caption{Minimum SNR versus the number of users for different schemes with $P=20$ dBm, $D=20$ m, and $N=\{4,8\}$.}
	\label{compare_K}
\end{figure}

Fig.~\ref{compare_K} depicts the minimum received \ac{SNR} versus the number of \acp{UE} for different schemes, where $P = 20$~dBm, $D = 20$~m, and $N = \{4, 8\}$. 
The Con-PASS and proposed \ac{2D-PASS} employ the continuous-position design. 
As expected, increasing the number of \acp{UE} leads to a gradual decline in the minimum \ac{SNR} of all schemes, primarily due to the power division among more \acp{UE}. 
Nevertheless, increasing the number of \acp{PA} provides additional spatial \ac{DoFs}, thereby improving the system’s beamforming capability and overall performance. 
For a fixed $N$, the proposed \ac{2D-PASS} consistently demonstrates higher \ac{SNR} values than both the conventional \ac{PASS} and the \ac{FPA}. 
For example, when $N = 4$ and $K = 6$, the \ac{2D-PASS} achieves about $1.3$~dB and $3.2$~dB gains over the conventional \ac{PASS} and \ac{FPA}, respectively. 
These findings highlight that the enhanced spatial flexibility and richer \ac{2D} configuration of the proposed \ac{PASS} architecture are particularly advantageous in supporting densely distributed multi-user environments.

\begin{figure}[t]
	\centering
	\includegraphics[width=0.65\linewidth]{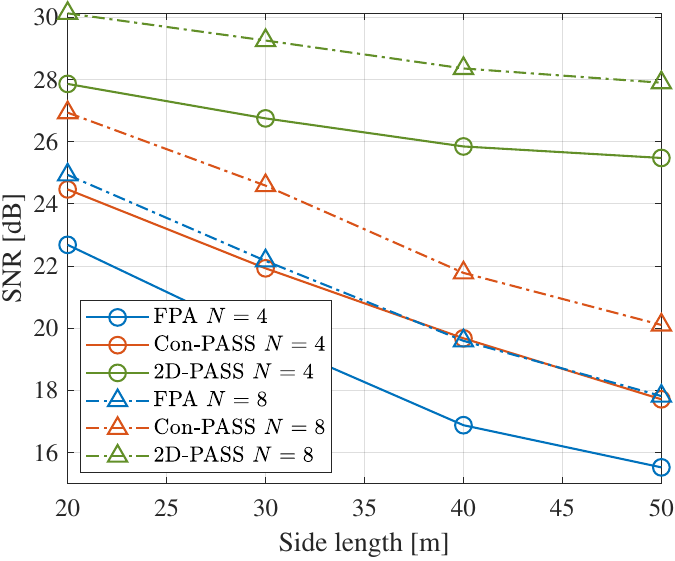}
	\caption{Minimum SNR versus side length for different schemes with $P=20$ dBm, $K=4$, and $N=\{4,8\}$.}
	\label{compare_D}
\end{figure}

Fig.~\ref{compare_D} shows the impact of the side length $D$ on the minimum received \ac{SNR} for different schemes, where $P = 20$~dBm, $K = 4$, and $N = \{4, 8\}$. 
Both the Con-PASS and the proposed \ac{2D-PASS} adopt the continuous-position design. 
As the coverage area expands, the received \ac{SNR} of all schemes gradually decreases due to the increased propagation distance and corresponding path loss. 
Meanwhile, increasing the number of \acp{PA} significantly enhances performance, as the additional spatial \ac{DoFs} facilitate more effective beamforming. 
For a fixed $N$, the proposed \ac{2D-PASS} consistently surpasses the conventional \ac{PASS} and the \ac{FPA} in performance. 
For instance, when $N = 4$ and $D = 40$~m, the \ac{2D-PASS} achieves approximately $6.1$~dB and $9$~dB \ac{SNR} gains compared with the conventional \ac{PASS} and \ac{FPA}, respectively. 
Moreover, when $D$ increases from $20$~m to $40$~m, the \ac{SNR}$\,$reduction of the \ac{2D-PASS} is limited to about $2$~dB, whereas the conventional \ac{PASS} and \ac{FPA} experience $4.6$~dB and $6$~dB drops, respectively. 
These results demonstrate the strong robustness of the proposed \ac{2D-PASS}, which effectively mitigates the impact of large-scale path loss through adaptive \ac{2D} spatial reconfiguration.

\begin{figure}[t]
	\centering
	\includegraphics[width=0.65\linewidth]{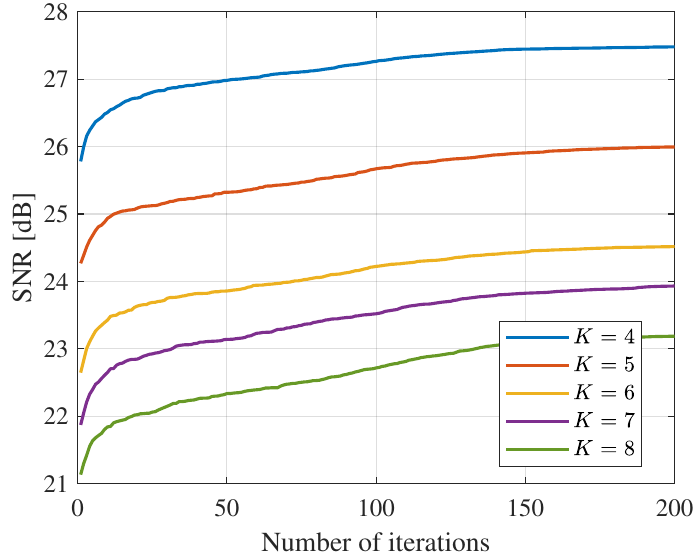}
	\caption{Convergence behavior of the proposed \ac{PSO}-based algorithm for different numbers of users with $P=20$~dBm, $D=20$~m, and $N=4$.}
	\label{compare_iterations}
\end{figure}

Furthermore, Fig.~\ref{compare_iterations} illustrates the convergence characteristics of the proposed \ac{PSO}-based optimization algorithm under different numbers of \acp{UE}, where $P = 20$~dBm, $D = 20$~m, and $N = 4$. 
As the number of iterations increases, the minimum received \ac{SNR} consistently improves and gradually stabilizes, indicating reliable convergence behavior. 
It is also evident that scenarios with fewer \acp{UE} achieve higher steady-state \ac{SNR} values, since more transmit power can be allocated to each \ac{UE}. 
All tested configurations exhibit smooth and monotonic convergence without oscillations, confirming the robustness and numerical stability of the proposed algorithm. 
These results verify that the designed \ac{PSO}-based framework can efficiently obtain considerable \ac{PA} configurations within a limited number of iterations, demonstrating both effectiveness and computational efficiency.

\begin{figure}[t]
	\centering
	\includegraphics[width=0.65\linewidth]{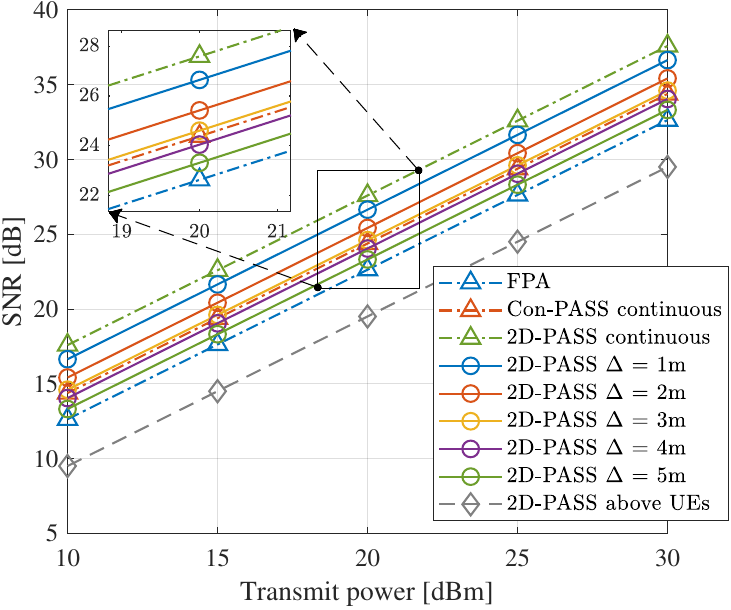}
	\caption{Minimum SNR versus transmit power for different continuous and discrete \ac{PASS} schemes with $K=4$, $D=20$ m, and $N=4$.}
	\label{compare_discrete}
\end{figure}

Subsequently, Fig.~\ref{compare_discrete} compares the minimum received \ac{SNR} of different positioning schemes for the discrete case, where $K = 4$, $D = 20$~m, and $N = 4$. 
The corresponding \ac{UE} distribution is illustrated in Fig.~\ref{compare_discrete}. 
It is evident that the continuous \ac{2D-PASS} achieves the highest \ac{SNR} across all cases, whereas the \ac{FPA} exhibits the lowest performance due to its fixed antenna placement. 
For discrete configurations, the achievable \ac{SNR} gradually decreases as $\Delta$ increases, since coarser quantization limits spatial adaptability and degrades beamforming precision. 
Nevertheless, at $P = 20$~dBm and $\Delta = 1$~m, the discrete \ac{2D-PASS} experiences only a marginal $1$~dB loss compared with its continuous counterpart. 
In addition, Fig.~\ref{compare_discrete} also presents the performance when the \acp{PA} are directly placed above the \acp{UE}. 
Although this configuration can partially reduce the path loss, the lack of phase optimization severely degrades the overall performance.
Given that discretization considerably relaxes hardware positioning accuracy and control complexity, this negligible degradation is well within practical limits, highlighting the feasibility and engineering viability of the proposed \ac{2D-PASS} design.

\begin{figure}[t]
	\centering
	\includegraphics[width=0.65\linewidth]{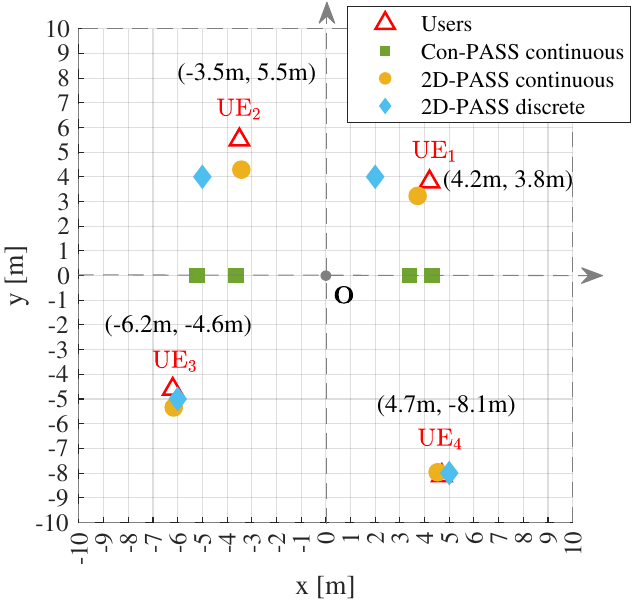}
	\caption{Optimized \ac{PA} positions for different schemes under a representative user distribution with $K=4$, $D=20$ m, and $N=4$.}
	\label{compare_positions}
\end{figure}

To provide a clearer visualization of the antenna adjustment results, Fig.~\ref{compare_positions} depicts the optimized \ac{PA} locations obtained by different schemes under a representative \ac{UE} distribution, where $K = 4$, $D = 20$~m, and $N = 4$. 
All configurations satisfy the minimum-spacing constraint in \eqref{op1:sub2}. 
It can be observed that the \acp{PA} are generally positioned as close as possible to the \acp{UE} within the feasible region, effectively mitigating path loss. 
In particular, each \ac{UE} is surrounded by at least one nearby \ac{PA}, which enhances the received signal strength and ensures the maximization of the minimum \ac{SNR} among all \acp{UE}. 
These findings clearly highlight the adaptive spatial reconfiguration capability of the proposed \ac{2D-PASS}, confirming its effectiveness in achieving user-centric analog beamforming.

\begin{figure}[t]
	\centering
	\includegraphics[width=0.65\linewidth]{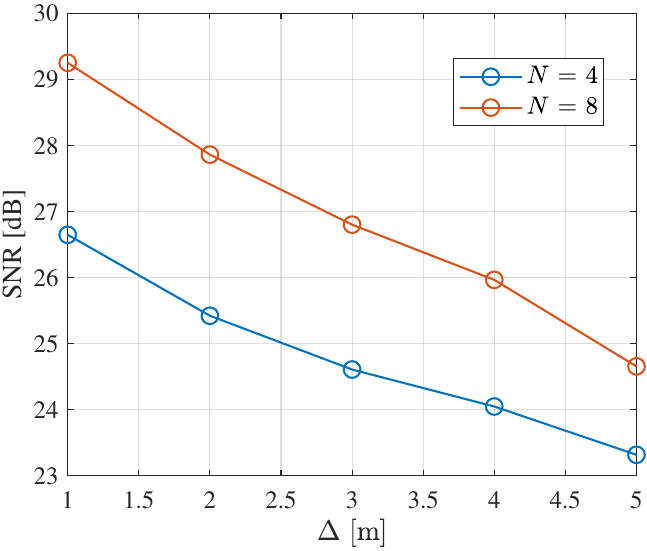}
	\caption{Minimum SNR versus position quantization step $\Delta$ for discrete \ac{2D-PASS} with $K=4$, $D=20$ m, and $N=\{4,8\}$.}
	\label{compare_Delta}
\end{figure}

Finally, Fig.~\ref{compare_Delta} presents the influence of the position quantization step $\Delta$ on the minimum \ac{SNR} performance for the discrete \ac{2D-PASS}, where $K = 4$, $D = 20$~m, and $N = \{4,8\}$. 
As expected, increasing $\Delta$ reduces the hardware implementation complexity but inevitably deteriorates the achievable \ac{SNR} due to the loss of spatial precision in antenna positioning. 
However, increasing the quantization resolution can yield significant performance gains. 
For instance, reducing the step size $ \Delta $ from $2\,\mathrm{m}$ to $1\,\mathrm{m}$ improves the minimum \ac{SNR} by approximately $4.8\%$.
In practical system design, an appropriate trade-off between performance and implementation cost can thus be achieved by selecting a suitable quantization step $\Delta$ in accordance with hardware constraints and precision requirements.

\section{Conclusion}\label{sec:conclusion}
 In this paper, we investigated a \ac{2D} extension of the conventional \ac{PASS}, termed 2D-PASS, which transforms the single-waveguide structure into an integrated dielectric plane. 
 A comprehensive analog beamforming optimization framework was developed to jointly configure the spatial positions of \acp{PA}, aiming to maximize the minimum SNR among all \acp{UE}.
 For the continuous configuration scenario, a \ac{PSO}-based approach was designed to effectively navigate the nonconvex search space, while a discrete version was further proposed to accommodate practical hardware constraints and finite \ac{PA} placement granularity. 
 The simulation results demonstrated that the proposed \ac{2D-PASS} achieved notable performance gains over conventional \ac{PASS} and \ac{FPA} schemes, particularly in multi-user and long-distance communication scenarios. 
Moreover, the \ac{2D} architecture enhances the spatial \ac{DoFs} for beam adaptation and exhibits improved robustness against large-scale path loss. 
These findings confirm the feasibility and advantages of employing \ac{2D-PASS} as a promising paradigm for flexible and energy-efficient antenna architectures in next-generation wireless systems.
 Future work may explore the integration of learning-based optimization and the experimental implementation of \ac{2D-PASS} for real-time adaptive beam control.
 


\ifCLASSOPTIONcaptionsoff
  \newpage
\fi

\bibliographystyle{IEEEtran}      
\bibliography{IEEEabrv,ref}

\end{document}